\documentstyle[prb,aps,epsf,twocolumn,eqsecnum]{revtex}
\begin{document}
\def\break#1{\pagebreak \vspace*{#1}}
\draft
\title{Scaling and criticality of the Kondo effect in a Luttinger liquid}
\author{Reinhold Egger and Andrei Komnik}
\address{Fakult\"at f\"ur Physik, Albert-Ludwigs-Universit\"at,
Hermann-Herder-Stra{\ss}e 3, D-79104 Freiburg, Germany}
\date{submitted to Physical Review B}
\maketitle
\widetext
\begin{abstract}
A quantum Monte Carlo simulation method has been developed and applied 
to study the critical behavior of a single Kondo impurity in a
Luttinger liquid. This numerically exact
method has no finite-size limitations and allows to simulate the
whole temperature range.  Focusing on the 
impurity magnetic susceptibility, we determine the 
scaling functions, in particular  for temperatures
well below the Kondo temperature. 
In the absence of elastic
potential scattering, we find Fermi-liquid behavior for 
strong electron-electron interactions, $g_c < 1/2$,
and anomalous power laws for $1/2<g_c < 1$, where $g_c$ is the
correlation parameter of the Luttinger liquid.
These findings resolve a recent controversy.
If elastic potential scattering is present, we find a logarithmically
divergent impurity susceptibility at $g_c<1/2$ which can be rationalized
in terms of the two-channel Kondo model.  
\end{abstract}
\narrowtext

\section{Introduction} 

Since its discovery, the Kondo problem
is one of the central topics in condensed-matter physics.\cite{kondo,hewson} 
It describes a magnetic spin-$\frac12$
impurity embedded into a metal and may be the simplest example for the 
growth of an effective
coupling at low energies, resulting in a nonperturbative ground state. For
normal metals this
ground state is found to be of Fermi-liquid type, where the 
quasi-particle wave functions simply acquire a phase shift.\cite{noz} 
The situation might change in
one-dimensional (1D) systems which are known to exhibit non-Fermi liquid 
behavior for arbitrary Coulomb interactions.  
The fundamental theory of interacting 1D metals in
the low-energy regime is the Luttinger liquid 
model.\cite{haldane,voit,schulz96} 
It is therefore of interest to understand the Kondo effect in a
Luttinger liquid.
An additional motivation arises from recent advances in nanofabrication which
now allow for controlled experiments on 1D systems.\cite{tarucha} 
In the future the question of how magnetic impurities behave when coupled to
1D metals might be of
crucial importance for experiments on quantum wires,\cite{tarucha,calleja}
carbon nanotubes,\cite{nature} or for edge states in the fractional quantum
Hall regime.\cite{chang}

The Luttinger liquid model unifies the
low-tem\-perature physics of many microscopic
lattice models for strongly correlated fermions, with only very few
phenomenological parameters. In particular, one has the dimensionless Coulomb
interaction strength parameters $g_c$ and $g_s$ for charge and spin sectors,
respectively, and the charge- and spin-density velocities $v_c$ and $v_s$.
The crucial assumptions are the absence of lattice instabilities
(like Umklapp scattering),
the absence of electron-electron backscattering, and 
that the Coulomb interaction potential is screened by mobile charge
carriers close to the 1D metal.
\break{1.6in}
As a simple model for interacting fermions, the Luttinger liquid model is
widely  used to study the 
influence of electronic correlations on dynamical properties of 1D 
metals, in particular in the presence of impurities.
 The case of a
spinless impurity is by now well understood.\cite{kane}
If the impurity has internal degrees of freedom, the
situation is more complicated and subject of this paper.

A Kondo impurity coupled to a Luttinger liquid was first considered by
Lee und Toner.\cite{toner} Employing the perturbative renormalization
group they established how the Kondo temperature $T_K$ depends on
the exchange coupling constant $J$. This turns out to be a power-law 
dependence,  while for normal metals $T_K$ is an exponential function of the
coupling constant.\cite{hewson} 
The same power law was found by Furusaki and Nagaosa.\cite{furu} These authors
derived the correct $SU(2)$ invariant scaling equations in the weak-coupling
regime and tentatively 
extended them to the strong-coupling regime, where a stable  
strong-coupling fixed point was found for both antiferromagnetic and
ferromagnetic exchange couplings. This strong-coupling fixed point describes a
many-body singlet formed by the impurity spin and the conduction electrons,
similar to what happens in a normal metal. 
Moreover, Furusaki and Nagaosa made detailed predictions concerning the
low-temperature critical properties of the impurity, e.g., the magnetic
susceptibility, the heat capacity, and the conductance. 
These quantities were found
to exhibit power-law behavior with interaction-dependent exponents.

However, it remained unclear whether the extra\-polation of the perturbative
scaling equations into the strong-coupling regime is justified.
Recent boundary conformal field theory (CFT) results by Fr\"{o}jdh and
Johannesson\cite{henrik} allow only two possible scenarios. Either the system
belongs to the Fermi liquid universality class or it indeed has the
properties predicted by Furusaki and Nagaosa. 
CFT itself is however unable to unambiguously decide which universality class
is ultimately realized for the Kondo problem in a Luttinger liquid. Several
recent papers seem to favor the local Fermi-liquid picture. Schiller and
Ingersent have discussed a truncated but related model which exhibits
Fermi-liquid behavior.\cite{schiller} In addition, according to the numerical
density-matrix renormalization group (DMRG) calculation of Wang,\cite{xwang}
Fermi-liquid behavior holds for a
spin-$\frac12$ impurity interacting with a 1D Hubbard chain. 
Recently, Chen {\em et al.}~deduced Fermi-liquid laws from the parity
and spin-rotation symmetry of a related model.\cite{luyu}
Here we shall address and resolve this controversial issue.

So far few studies have dealt with magnetic impurities exhibiting
elastic potential scattering in addition to the conventional (Kondo) exchange
coupling.\cite{wang,balseiro}  Why it should be considered at all 
becomes clear after the following discussion.
If one starts out from the usual Anderson model
to describe a localized orbital interacting with conduction electrons,
the natural generalization in 1D would include Coulomb interactions
among the conduction electrons.\cite{schork}
For uncorrelated conduction electrons,
one can then derive the usual Kondo exchange coupling in the local moment 
regime [which is realized for large on-site repulsion 
and a single-particle impurity level deep below the Fermi energy] 
by applying the Schrieffer-Wolff transformation.\cite{hewson}
This transformation generates the exchange coupling [see Eq.~(\ref{contact})
below] and, in addition, an elastic potential
scattering term. In the correlated case of interest here, this latter term
may be crucial since elastic potential scattering is relevant in a Luttinger
liquid.\cite{kane} 
What  can happen to a magnetic impurity  in a Luttinger
liquid  in the presence of strong potential scattering was first studied
by Fabrizio and Gogolin.\cite{fabrizio} They predict that 
at the strong-coupling point for the Kondo effect discussed
above elastic potential scattering is
irrelevant for rather weak repulsive Coulomb interaction, namely for 
$1/2<g_c<1$. In contrast, for strong enough interaction, $g_c<1/2$,  
the potential scattering breaks up the system into two independent
chains. The magnetic impurity then interacts with two subsystems (channels),
and the two-channel Kondo picture emerges.\cite{fabrizio}  

In this paper we present a 
path-integral quantum Monte Carlo (QMC) method allowing for the
computation of thermodynamic properties of a Kondo impurity in a Luttinger
liquid. This method is numerically exact within the statistical
error bars inherent to the MC technique.
The main advantages of our method are the absence of any system-size
restriction, contrary to DMRG simulations or several QMC lattice
algorithms,\cite{binder} and the possibility
of treating arbitrarily correlated conduction electrons. 
However, as is well known, QMC simulations of spin systems often have 
to deal with the
fundamental sign problem.\cite{loh,mak}
It is caused by sign alternations of the QMC weight function for different
system configurations. This generally leads to a small
signal-to-noise ratio and hence to numerical instabilities. 
The approach we shall discuss here is also plagued by a 
sign problem. However, our sampling technique moderates the
problem to an extent allowing us to treat 
sufficiently low temperatures.\cite{foot} 
Moreover, we have also applied filtering
techniques\cite{mak} which provide a general method to ease
the mentioned difficulties. Yet the approach described below suffers
only from a minor intrinsic sign problem, and the use of the
filtering technique 
is not really crucial to obtain the results described below. 
Finally we mention that for a calculation 
of the Kondo screening cloud around the impurity, a simpler
version of the present simulation method has 
been employed by Egger and Schoeller.\cite{egger}

One might ask why we chose to develop a new algorithm 
for studying the Kondo effect even though the exceptionally stable
and widely used QMC impurity algorithm due to Hirsch and 
Fye \cite{hirsch} is available.  The reason is that we
have to include the Coulomb interactions
among the conduction electrons which are responsible for the Luttinger 
liquid state. In the Hirsch-Fye algorithm, one 
traces out the conduction electron degrees of freedom away
from the impurity and then updates only the arising fermion
determinant.  However, by construction, this procedure works
only if the conduction electrons are in the Fermi liquid state.
By employing the bosonization method, as is shown below, one 
can in fact follow a similar route as in Ref.\onlinecite{hirsch}
and trace out the now correlated conduction electrons away from
the impurity. 
From a historical point of view, it may perhaps seem surprising that 
the QMC technique has not been a major tool
in the resolution of the conventional Kondo problem 
where conduction electrons are described by a Fermi liquid.
As realized in Ref.\onlinecite{schotte}, the main obstacle
is the exponentially small Kondo temperature,  which in turn
requires the study of extremely
low temperatures difficult to achieve in path-integral MC
calculations.  For the Kondo effect in a Luttinger liquid,
however, the Kondo temperature is much higher and QMC simulations
become feasible even in the asymptotic low-temperature regime.

The outline of this paper is as follows. In Sec.~II we
discuss the Luttinger liquid model with a Kondo impurity
and describe our Monte Carlo algorithm in some detail.
In Sec.~III results for a Kondo impurity in the absence
of elastic potential scattering are presented, and Sec.~IV
gives results in the presence of additional strong elastic
potential scattering. Finally, some concluding remarks
are offered in Sec.~V.

\section{Theory and Quantum Monte Carlo Method}

The low-energy properties of correlated 1D systems are
most conveniently described in terms of the bosonization
method.\cite{haldane,voit,schulz96} 
The spin-$\frac12$ electron field operator is
expressed in terms of spin and charge boson fields
which obey the algebra (we put $\hbar=1$)
\begin{equation}
[\theta_i(x),\varphi_j(x')] = -\frac{i}{2}\delta_{ij} {\rm sgn}(x-x')\;,
\end{equation}
where $i,j$ denote the charge ({\em c}) or spin ({\em s}) degrees of freedom. 
The canonical momentum for the $\varphi_i$ phase field is
therefore $\Pi_i(x)=\partial_x\varphi_i(x)$. The two kinds
of phase fields are not independent but basically dual
fields.  Written in terms of the boson phase fields, the right- or 
left-moving ($p=\pm$) component of the 
electron annihilation operator for spin $\sigma=\pm$
takes the form
\begin{eqnarray}\nonumber
\psi_{p\sigma}(x) &=& \sqrt{\frac{\omega_c}{2\pi v_F}} \,
\eta_{p\sigma} \exp\left[-i\sqrt{\pi/2} [\theta_c(x)+\sigma
\theta_s(x)] \right] \\
\label{psidef}
&\times& \exp\left[ipk_F x + ip\sqrt{\pi/2} [\varphi_c(x)+
\sigma \varphi_s(x)] \right] \;.
\end{eqnarray}
The bandwidth cutoff is $\omega_c$,
and we put $\omega_c=v_F k_F=1$
in what follows ($v_F$ is the Fermi velocity). A corresponding
lattice constant  can then be 
defined as $a=v_F/\omega_c$. 
In Eq.~(\ref{psidef}), we have also included real Majorana fermions
$\eta_{p\sigma}$. Their purpose is to ensure proper anticommutation
relations between operators for different branches labeled
by $p\sigma$. Since only products $\eta_{p\sigma}\eta_{\pm p \pm \sigma}$
will appear in the Hamiltonian, a convenient choice for these products is (see
also Ref.~\onlinecite{egger})
\begin{eqnarray} \nonumber
\eta_{p\sigma} \eta_{-p\sigma} & = & ip \sigma {\tau}_z   \; , \\
\eta_{p\sigma} \eta_{p-\sigma} & = & i\sigma {\tau}_y
\label{maj}  \; , \\ \nonumber
\eta_{p\sigma} \eta_{-p-\sigma} & = &i p {\tau}_x \;,
\end{eqnarray}
where ${\tau}_i$ are the usual Pauli matrices. There is a simple way
to realize why Eq.~(\ref{maj}) holds. Keeping in mind that
$\eta_{p\sigma}\eta_{p\sigma}=1$ for all $p$ and $\sigma$, one can easily
check that  all products of the operators $\eta_{p\sigma}\eta_{p'\sigma'}$ 
defined
in Eq.~(\ref{maj}) fulfill the correct algebra required by anticommutation
relations for the $\eta_{p\sigma}$. 
Actually, Eq.~(\ref{maj}) shows only
one of several possibilities to choose representations of Majorana fermion 
products. Of course, one can verify that the subsequent results do not depend
on which one we choose.

Under the conditions specified in the Introduction, the effective low-energy
Hamiltonian for the clean electronic system takes the simple
Gaussian form of the bosonized Luttinger liquid 
model,\cite{haldane,voit,schulz96}
\begin{equation} \label{mmm}
H_0=\sum_{j=c,s} \frac{v_F}{2} \int dx\,[\Pi_j^2 + g_j^{-2}
(\partial_x \varphi_j)^2 ]\;.
\end{equation}
In a system with full (Galilean) translation invariance, 
the velocities $v_c$ and
$v_s$ are required to 
fulfill $v_i = v_F/g_i$. 
We have assumed this relation in Eq.~(\ref{mmm}), bearing in mind
that  for lattice models it need not be fulfilled.\cite{schulz96} 
A general rule of thumb for the dimensionless interaction strength parameter
$g_c$ is 
\[
 g_c \approx [ 1 + 2U/\pi v_F]^{-1/2} \; ,
\]
where $U$ is the forward-scattering amplitude of the screened Coulomb
interaction. 
In the important case of repulsive interactions, $g_c<1$. The spin parameter
should be set to $g_s=1$ in
order to respect the underlying spin isotropy of the electrons.\cite{schulz96}
This value is also the fixed point value of the renormalization group (RG) if 
one incorporates electron-electron backscattering.  In the remainder, 
we shall put $g_s=1$ and neglect backscattering. Following
the usual perturbative RG analysis, this
could at most lead to weak logarithmic corrections\cite{voit} 
to the power laws found below.

Next we consider what happens once a single magnetic impurity is brought into
the Luttinger liquid, say, at $x=0$. We envision
a spin-$\frac12$ impurity characterized
by the spin operator ${\vec{S}}=\frac12{\vec{\tau}}$, where 
${\vec{\tau}}$ denotes the vector of Pauli matrices [this is not to be
confused with the ${\tau}_i$ appearing in Eq.~(\ref{maj})]. 
In terms of the conduction electron spin density operator 
\begin{equation} \label{spdens}
\ {\vec{s}}(x) = \frac12 \sum_{pp'\sigma\sigma'}
\psi^\dagger_{p\sigma}\,  {\vec{\tau}}_{\sigma\sigma'}
 \psi_{p'\sigma'}
\end{equation}
and a point-like exchange coupling $J$, the standard contact
contribution to the Hamiltonian reads with ${s}_{\pm}=\ {s}_x\pm
i {s}_y$
\begin{equation} \label{contact}
H_I=J{\vec{s}}(0){\vec{S}} = J{s}_z(0){S}_z + \frac{J}{2} \left(
 {s}_+(0){S}_- +{s}_-(0){S}_+ \right) \;.
\end{equation}
We consider only antiferromagnetic values $J>0$ in this paper. 
Using the bosonization formula (\ref{psidef}), 
the spin density (\ref{spdens}) of the conduction electrons reads
 \begin{eqnarray} \nonumber
 {s}_z(x)& =&  \frac{1}{\sqrt{2\pi}} \,\partial_x \varphi_s(x)\\
\nonumber &+&
\frac{1}{\pi a}{\tau}_z \sin[2k_Fx + \sqrt{2\pi}\,\varphi_c(x)]
\cos[\sqrt{2\pi}\,\varphi_s(x)] \\ 
 {s}_\pm(x) &=& \frac{1}{\pi a} \exp[ \pm\sqrt{2\pi}\,i\theta_s (x) ]
\Bigl\{\pm i {\tau}_y \cos[\sqrt{2\pi}\,\varphi_s(x)]  \nonumber \\
 \label{spindens}
&+&{\tau}_x \sin[2k_F 
x+\sqrt{2\pi}\,\varphi_c(x)]\Bigr\}\;.
\end{eqnarray}
Here the ${\tau}_i$ matrices  come from the Majorana fermion products
(\ref{maj}). 

Now we can incorporate elastic potential scattering. For that purpose, we need
the bozonized form of the total electron density operator, 
\begin{eqnarray} \label{densop}
\rho(x) &=& \sqrt{2/\pi} \, \partial_x \varphi_c(x) \\ \nonumber
&+& \frac{2}{\pi a}{\tau}_z \cos[2k_F x + \sqrt{2\pi} \varphi_c(x)]
\sin[\sqrt{2\pi} \varphi_s(x)] \;,
\end{eqnarray}
where we have omitted the background charge density $2k_F/\pi$.
The $2k_F$ component stems from terms mixing right- and left-moving
particles, while the slow component $\sim \partial_x
\varphi_c$ comes directly from the densities of right- and left-movers.
There is also a $4k_F$ component in $\rho(x)$ not specified in 
Eq.~(\ref{densop}) which dominates in the regime $g_c<1/3$.
Since in that limit the Coulomb interactions are extremely
strong, any elastic potential scattering will be highly
relevant.  Including a point-like scattering potential of strength $V$, one
obtains first a forward-scattering contribution 
$H_{FS} = V \sqrt{2/\pi}\partial_x \varphi_c(0)$.
This can simply be absorbed by a phase shift
in the $\sin[2k_F x]$ or $\cos[2k_F x]$ factors and is 
therefore omitted in the sequel. We are then left with the 
important backscattering contribution
\begin{equation}\label{hv}
H_V= \frac{2V}{\pi v_F} \tau_z \cos[\sqrt{2\pi} \varphi_c(0)]
\sin[\sqrt{2\pi} \varphi_s(0)] \;.
\end{equation}

For numerical calculations, it is advantageous to
employ a unitarily transformed picture such that
the Hamiltonian becomes explicitly real-valued. This is achieved by 
choosing \cite{egger,kivelson}
\begin{equation}\label{u1}
U = \exp[\sqrt{2\pi} i \theta_s(0) S_z]\;,
\end{equation}
such that 
\begin{eqnarray} \nonumber 
 U s_\pm S_\mp U^\dagger 
&=& \frac{1}{\pi a} \Bigl \{ \pm i \tau_y \cos[\sqrt{2 \pi}\varphi_s(x)]  \\ 
 &+& \tau_x \sin[2 k_F x + \sqrt{2 \pi}\varphi_c(x)] \Bigr \} S_\mp \; . 
\end{eqnarray}
The total transformed Hamiltonian $\widetilde{H}=U^{}HU^\dagger$
then reads
\begin{eqnarray} \label{h1}
 \widetilde{H} & = & H_0 + \frac{1}{\sqrt{2\pi}} \bar{J} S_z 
\partial_x \varphi_s(0)\\ \nonumber
&+&\frac{J}{\pi v_F} 
 \Bigl( \tau_x S_x \sin[\sqrt{2\pi}\varphi_c(0)]   
+ \tau_y S_y \cos[\sqrt{2\pi}\varphi_s(0)]
\\ \nonumber
&+& \tau_z S_z \sin[\sqrt{2\pi} 
\varphi_c(0)] \cos[\sqrt{2\pi} \varphi_s(0)] \Bigr) \\ \nonumber
&+& \frac{2V}{\pi v_F} \tau_z \cos[\sqrt{2 \pi} \varphi_c(0)]
\sin[\sqrt{2 \pi} \varphi_s(0)] \;,
\end{eqnarray}
where the transformation leads to a change in the forward scattering,
\begin{equation}
\bar{J} = J - 2\pi v_F\;.
\end{equation} 
The unitary transformation (\ref{u1}) also removes the 
$\theta_s(0)$ field from the Hamiltonian (in fact, it is
just constructed to remove this phase factor). 

From Eq.~(\ref{h1})
it is obvious that the Majorana fermions are dynamically constrained
to follow the impurity spin dynamics since 
\begin{equation} \label{grund}
 [{\tau}_z \otimes{S}_z, \widetilde{H} ] = 0 \; .
\end{equation}
Therefore we must have 
\begin{equation} \label{cons}
 \tau_z = \pm \,2 S_z \;.
\end{equation}
The only manifestation of the Majorana fermions is 
the overall sign, which we set equal to $+1$ in the following. 

One can simplify the total Hamiltonian using 
properties of the products $\tau_k \otimes{S}_k$ appearing in
Eq.~(\ref{h1}). Evaluated in the 
$|S_z, \tau_z\rangle$ basis, we find from Eq.~(\ref{cons}) 
\begin{eqnarray} \nonumber
 \langle S_z'\tau_z'|{S}_x \otimes{\tau}_x| S_z \tau_z \rangle
&=& \frac{1}{2}  \delta(S_z, -S_z') = 
\langle S_z'|{S}_x | S_z \rangle \; ,  \\
\nonumber \langle S_z'\tau_z'|{S}_y \otimes{\tau}_y| S_z \tau_z \rangle
&=& - \frac{1}{2} \delta(S_z,
-S_z') = - \langle S_z'|{S}_x | S_z \rangle \; ,  \\ 
\label{matrixelements}
 \langle S_z'\tau_z'|{S}_z \otimes{\tau}_z| S_z \tau_z \rangle
&=& \frac{1}{2} \delta(S_z, S_z') = \frac{1}{2}\langle S_z'| S_z \rangle \; .
\end{eqnarray}
Therefore we can reduce the original Hamiltonian through the substitutions
\begin{eqnarray}\nonumber
{S}_x \otimes{\tau}_x &\rightarrow&{S}_x \; , \\
{S}_y \otimes{\tau}_y &\rightarrow& -{S}_x \; ,  \\ \nonumber
{S}_z \otimes{\tau}_z &\rightarrow& 1/2  \; . 
\end{eqnarray}
This leads from Eq.~(\ref{h1}) to
\begin{eqnarray} \nonumber 
 \widetilde{H} &=& H_0 + \frac{1}{\sqrt{2 \pi}} \bar{J}{S}_z \partial_x
\varphi_s(0) \\  \nonumber  
 &+& \frac{J}{2\pi v_F} \left\{ 2{S}_x \left( \cos[ \sqrt{2 \pi}
\varphi_c(0)]  \right. \right.  - \left.
\cos[ \sqrt{2 \pi} \varphi_s(0)] \right) \\ \nonumber
&+& \left.  \cos[ \sqrt{2 \pi}
\varphi_c(0)] \cos[ \sqrt{2 \pi} \varphi_s(0) ] \right\} \\
\label{fla}
 &-& \frac{2V}{\pi v_F} 2{S}_z \sin[ \sqrt{2 \pi} \varphi_c(0)]
\sin[ \sqrt{2 \pi}\varphi_s(0)] \; .
\end{eqnarray}
For further convenience, we have shifted the $\varphi_c$ field by 
$\sqrt{\pi/8}$. This changes
$\sin[ \sqrt{2 \pi}\varphi_c]$ to $\cos[ \sqrt{2 \pi}\varphi_c]$ and 
$\cos[ \sqrt{2 \pi} \varphi_c]$ to $-\sin[\sqrt{2 \pi}\varphi_c]$. 

We  now proceed by integrating out
all boson fields $\varphi_j(x)$ for $x\neq 0$ for a given impurity spin path,
as these represent just Gaussian integrations.  
The Euclidean action can then be 
expressed as an average over new fields ($t$ denotes Euclidean time
extending from $t=0$ to $t=\beta=1/k_B T$) 
\begin{equation}
q_j(t) = \sqrt{2\pi} \varphi_j(x=0,t)\;,
\end{equation}
with the constraint being enforced by Lagrange multiplier fields
$\lambda_j(t)$. Since the spin and charge modes are
only coupled through the terms $\sim J$ and $\sim V$ in  Eq.~(\ref{fla}),
the elimination of the $\varphi_j$ degrees of freedoms
can be carried out independently for $j=c$ and $j=s$.
As the computation for the charge part follows
the same line of reasoning as for the spin part
(it can be obtained by retaining
$g_c$ factors and disregarding the $\sim \bar{J}$ term), we
only discuss the elimination of the $\varphi_s$ field in the following.
Results for the $c$ field are then recovered at the end, 
see Eq.~(\ref{seff}). 

After a partial integration, we have to integrate out $\varphi_s(x,t)$  
from a problem characterized by the effective  action 
\begin{eqnarray*}
S_{\rm eff} &=&\frac{1}{2v_F} \int dx dt
\left[ (\partial_t \varphi_s)^2 + v_F^2 (\partial_x \varphi_s)^2 \right]
\\ &-& \frac{\bar{J}}{\sqrt{2\pi}} \int dx dt \,
 \varphi_s(x,t) S_z(t) \delta^\prime(x) \\ &+& i\int dt\, \lambda_s(t) 
[q_s(t)-\sqrt{2\pi} \varphi_s(0,t)]\;.
\end{eqnarray*}
This can be achieved by solving the Euler-Lagrange equation
\begin{eqnarray*}
&& (\partial_t^2 + v_F^2 \partial_x^2) \varphi_s(x,t) =
\nonumber \\
&& \qquad - \sqrt{2\pi} i \Bigl[\lambda_s(t)
\delta(x) - i \frac{\bar{J}}{2\pi} S_z(t) \delta^\prime(x) \Bigr]\;,
\end{eqnarray*}
which is easily done in Fourier space,
\begin{equation}
\varphi_s(x,t) = \frac{1}{\beta} \sum_{n=-\infty}^\infty \;
\int_{-\infty}^\infty 
\frac{dk}{2\pi} \, e^{i\omega_n t+ ikx} \varphi_s(k,\omega_n)\;,
\end{equation}
with similar relations for other fields ($\omega_n=2\pi n/\beta$
are the Matsubara frequencies). 
Inserting the solution of the Euler-Lagrange equation for $\varphi_s$
into $S_{\rm eff}$, one has
\begin{eqnarray*}
S_{\rm eff} &=&  \frac{i}{\beta} \sum_n q_s(\omega_n) 
\lambda_s(-\omega_n) \\ &+& \frac{1}{2\beta}\sum_n \Bigl[
\lambda_s(\omega_n) \lambda_s(-\omega_n) F_s(0,\omega_n)\\
&+& (\bar{J}/2\pi)^2 F_s^{\prime\prime}(0,\omega_n) S_z(\omega_n)
S_z(-\omega_n)\Bigr]\;.
\end{eqnarray*}
Here we have defined the boson propagators\cite{grabert} 
($j=c,s$)
\begin{eqnarray}\label{bosprop}
F_j(x,\omega)  &=& v_F \int_{-\infty}^\infty dk \,\frac{\exp(ikx)}{\omega^2
+ v_j^2 k^2}  \\  \nonumber
&=& \frac{\pi g_j}{|\omega|} \exp(-|\omega x/v_j|) \\ \nonumber  
F^{\prime\prime}_j(x,\omega) &=& (\partial^2/\partial x^2) F_j(x,\omega)
\\ \nonumber &=&
-\frac{2\pi g_j}{v_j} \delta( x) + \frac{\pi g_j |\omega|}{v_j^2}
\exp(-|\omega x/v_j|)\;. 
\end{eqnarray}
The $\delta(x)$-contribution to $F^{\prime\prime}_j(x,\omega)$ is
irrelevant in our case, since it causes only a constant term 
$\sim \int dt S_z^2(t)$ in the effective action.
We therefore disregard it in the following. Finally, the 
Lagrange multiplier field can be integrated out by simple minimization,
\begin{equation}
\lambda_s(\omega_n) = - i \frac{q_s(\omega_n)}{F_s(0,\omega_n)}\;.
\end{equation}
Collecting results, the effective action is found to read
\begin{eqnarray} \label{seff}
S_{\rm eff} & = & \sum_{j=c,s} \frac{1}{2\pi g_j \beta } \sum_n
|\omega_n| |q_j(\omega_n)|^2 \\ \nonumber 
&+& \frac{J}{2\pi v_F}\int \, dt \,\cos[q_c(t)]\cos[q_s(t)]  \\ \nonumber
&-& \frac{2 V}{\pi v_F} \int \, dt \, 2S_z(t) \sin[q_c(t)]\sin[q_s(t)] 
\\  &+& \frac{\bar{J}^2}{8\pi\beta} \sum_n
|\omega_n| |S_z(\omega_n)|^2 + S' \;, 
\nonumber
\end{eqnarray}
with $S'$ formally given as $\int dt H'(t)$ with 
\begin{equation} \label{hjj}
H'(t)  = \frac{J}{2\pi v_F} 2S_x(t) \Bigl\{\cos[q_c(t)]
- \cos[q_s(t)]\Bigr\}\; . 
\end{equation}

After these preparations, we now proceed further and describe a  
quantum Monte Carlo (QMC) algorithm for this problem. 
Since the unitary transformation $U$ given in Eq.~(\ref{u1}) leads
to the real-valued Hamiltonian (\ref{fla}), it is very convenient to
employ this representation. 
We have focused on the impurity susceptibility 
\begin{equation} \label{defform}
 \chi = \int_0^\beta \, dt \, \langle S_z(t)S_z(0) \rangle \; , 
\end{equation}
since knowledge of $\chi$ at low temperatures is sufficient to answer
the questions raised in the Introduction. 
Here the average is taken using Eq.~(\ref{fla}). 
The impurity spin operator $S_z$ does not change
under the unitary transformation $U$, so the expression (\ref{defform}) holds
also in the transformed picture.

The QMC simulation scheme starts out from the discretized imaginary-time
path-integral representation for Eq.~(\ref{defform}) 
 using the effective action (\ref{seff}) with
(\ref{hjj}). The imaginary-time
slice is $ \delta t= \beta/N$, where the Trotter number $N$ should
be large enough.  In practice, one has to
check empirically at the end that  results converge upon increasing $N$. 
In the QMC simulations, a hard cutoff was chosen by keeping only
Matsubara frequencies $|\omega_n| < \omega_c$. The sampling
of the $q_j$ fields is then most conveniently carried out directly using their
Matsubara components. For the impurity spin
variable, however, it is mandatory to use the time
representation because one has a discrete variable $S_j= 2S_z(t_j)=\pm 1$,
where $t_j=j\delta t$ is the $j$th time slice.
The action contribution $S'$ is now determined
as follows. From a Trotter breakup procedure\cite{binder}
valid at small enough $\delta t$, we obtain the representation
\begin{equation} \label{skappa}
\exp(-S') = \prod_{j=1}^N \langle S_{j+1}|
\exp[- \delta t\, H'(t_j) ] | S_j\rangle \; ,
\end{equation}
where the spins obey periodic boundary conditions, $S_{N+1}=S_1$.
Using the matrix elements (\ref{matrixelements}), we obtain 
(up to an irrelevant overall constant)
\begin{equation} \label{trotter1}
 \exp(-S') = \prod_{j=1}^{N} \left\{ e^{f(t_j)} + S_jS_{j+1}e^{-f(t_j)}
\right\}  \; , 
\end{equation}
where
\begin{equation} \label{m}
 f(t) =  \frac{J \delta t}{2\pi v_F} \Bigl\{ \cos[q_c(t)] - 
\cos[q_s(t)] \Bigr\} \; .
\end{equation}

The QMC sampling is then drawn from the weight function
\begin{equation}\label{wwww}
 {\cal P} \sim | \exp(-S_{\rm eff}) |  \;,
\end{equation}
where $S_{\rm eff}$ is specified in Eq.~(\ref{seff}) together
with Eq.~(\ref{trotter1}). Since
$\exp(-S')$ can be negative, the simulations have
to face the sign problem.\cite{loh} 
For not exceedingly large
$J$ and low temperatures, however, the sign problem is not severe and the QMC
algorithm described here can be applied to a wide 
region of the parameter space without instabilities.  
Denoting the sign of the MC weight as 
\begin{equation}
\xi_p={\rm sgn}\Bigl\{ \exp(-S') \Bigr\} \;,
\end{equation}
the MC denominator will then be $\langle \xi_p \rangle$.
The severity of the sign problem  is usually measured in terms
of $\langle \xi_p \rangle$.\cite{loh} 
One way to weaken the sign problem is to employ the Mak filtering
technique\cite{mak} which can improve the stability of the
algorithm by about 20 to 30\%. For the results presented below, this technical
trick was not necessary, and good statistics can be acquired even without a
filtering method.

Of particular interest is the value of the impurity 
susceptibility $\chi$ which is given by the temperature-dependent expression
\begin{equation} \label{osnow}
 \chi =  \frac{\delta t}
{4}\frac{\langle \xi_p \sum_{j=1}^N S_j S_1 \rangle}{\langle \xi_p
\rangle} \; .
\end{equation}
Here the Monte Carlo sampling over the configuration space spanned by the
variables $\{q_c(\omega_n), q_s(\omega_n), S_j \}$ is carried out 
using the weight (\ref{wwww}).

\begin{figure}
\epsfysize=7.5cm
\epsffile{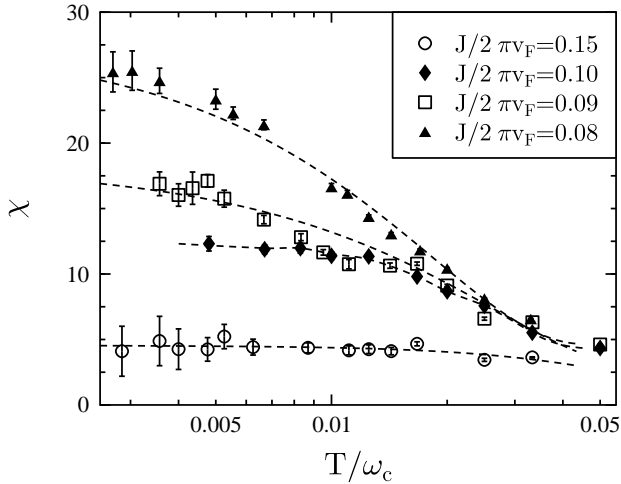}
\caption[]{\label{fig1}
Low-temperature behavior of the impurity magnetic susceptibility at
$g_c=1/4$ and
various values of the coupling constant $J$. Notice the semi-logarithmic
scales. Dashed curves represent guides to the eye only. Vertical bars give
standard deviation error bars due to the MC sampling.}
\end{figure}

For the results presented below, the average sign
is $\langle \xi_p \rangle \approx 0.1$, but in
 practice stable simulations can be carried out
even for $\langle \xi_p \rangle \approx 0.01$ at the expense of long central
processing unit (CPU) times. The Monte Carlo trajectory was drawn from the
standard Metropolis algorithm.\cite{binder}
 We have used local updates of the phase
fields at $x=0$, i.e., of the Matsubara components $q_c(\omega_n)$ and
$q_s(\omega_n)$  for $|\omega_n| \leq \omega_c$, and of the impurity spin
trajectory $S_z(t_j)=S_j/2= \pm 1/2$. Typical discretization parameters [for
$J/2 \pi v_F= 0.1$ and $V=0$] required to ensure convergence to the continuum
limit of the discretized path integral are $\omega_c  \delta t\simeq 0.3$. The
acceptance ratios for local updates of the $\{ S_j \}$ variables
are rather low for the
parameter values considered below, typically of the order of $5\%$.
Therefore data are accumulated only after at least $5$ full MC passes to
ensure statistical independence. Our code performs at an average speed of 1
CPU hour per $5000$ samples (separated by $5$ MC passes) on an IBM RISC
6000/model 590 workstation at the lowest temperatures under consideration.
Results reported here require typically $10^6$ samples per data 
point.

\section{Critical impurity dynamics without potential scattering}

In this section we study the case without potential scattering $(V=0)$, with
particular emphasis on the controversy about the low-temperature scaling. 
From our QMC data we observe that  all 
$\chi(T)$ curves for different coupling constants $J$ but at a given Coulomb
interaction strength $g_c$ can be mapped onto a single universal scaling
curve.
\begin{figure}
\epsfysize=7.5cm
\epsffile{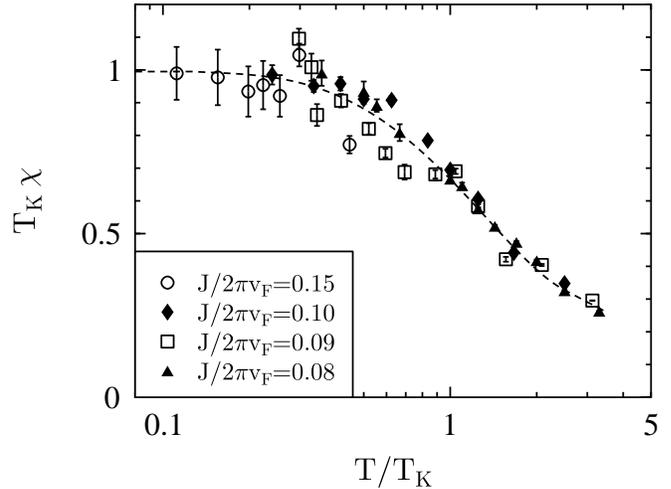}
\vspace{0.3cm}
\caption[]{ \label{fig2}
Scaling curve for Fig.~\ref{fig1}. Notice
the semi-logarithmic scales. The dashed curve is a guide to the eye only.
}
\end{figure}
 For instance, our raw
data for $g_c=1/4$ are shown in Fig.~\ref{fig1}, and the scaling curve for
$\chi$ is depicted in Fig.~\ref{fig2}. Apparently, there is a universal
scaling function $f$ such that 
\begin{equation} \label{TK1}
 T_K \, \chi(T) = f(T/T_K)  \; . 
\end{equation}
Using this matching procedure, the Kondo temperature can be determined
straightforwardly. On the other hand, the value $\chi_0=\chi(T=0)$ is finite
and can be used to define the Kondo temperature as well. Indeed, the
zero-temperature magnetic susceptibility 
is of the order of the binding energy of
the many-body singlet state formed by the impurity spin and the conduction
electrons. Therefore we can fix $T_K$ alternatively as
\begin{equation} \label{TK2}
 \chi_0 = 1/T_K \; , 
\end{equation}
which implies $f(0)=1$ from Eq.~(\ref{TK1}). From the zero-temperature limit
of $\chi(T)$ [which can be obtained quite accurately by extrapolation of the
data] we can then read off $T_K$.
By means of either of these two prescriptions, 
as is shown in Fig.~\ref{fig3} for
$g_c=1/4$, one can indeed verify the dependence
of the Kondo temperature on the coupling constant predicted in
Refs.~\onlinecite{toner} and \onlinecite{furu},  
\begin{equation}   \label{TK}
 T_K = D \left( \frac{J}{2 \pi v_F} \right)^{2/(1-g_c)} \; ,
\end{equation}
where $D$ is of the order of the bandwidth cutoff $\omega_c$.
Simultaneously, one gets the universal scaling curve for given interaction
strength $g_c$.

Now we wish to address the low-temperature critical behavior. The
low-temperature form ($T \ll T_K$) of the impurity susceptibility
(\ref{TK1}) can exhibit only two possibilities 
allowed from conformal field theory (CFT).\cite{henrik}
Either one has (i) Fermi-liquid behavior,
\begin{equation} \label{fermi}
  f(T/T_K) = 1 - c_1 (T/T_K)^2 + \cdots \; ,
\end{equation}
or (ii) the anomalous exponents predicted by Furusaki and Nagaosa,\cite{furu} 
 \begin{equation} \label{TL}
  f(T/T_K) = 1 - c_2 (T/T_K)^{1/g_c} + \cdots \;,
 \end{equation}
where $c_1$ and $c_2$ are positive constants.
Obviously, at $g_c=1/2$ one must see the $T^2$ behavior. This is a check for
our numerics which is indeed passed nicely. The results 
(taking $J/2 \pi v_F=0.1$) are presented in
Fig.~\ref{fig4}. In the inset we have depicted the dependence of the
deviation $(\chi_0-\chi) T_K$ on the thermal scaling variable
$T/T_K$. As one can see, the correct $T^2$ power law emerges, provided we are 
below the Kondo temperature.  

\begin{figure}
\epsfysize=7.5cm
\epsffile{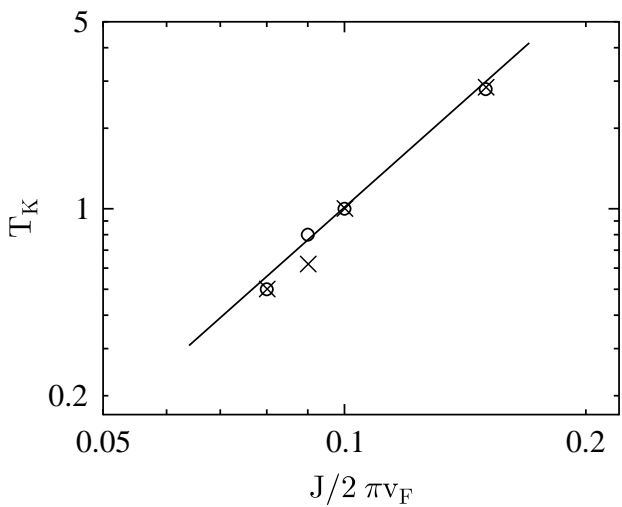}
\vspace{0.3cm}
\caption[]{\label{fig3}  Normalized
Kondo temperature $T_K$ as a function of the
exchange coupling $J$ for $g_c=1/4$.
The normalization has been chosen such that $T_K=1$ for $J/2\pi v_F = 0.1$.
The crosses represent the evaluation of $T_K$ from Eq.~(\ref{TK1})
and the circles from Eq.~(\ref{TK2}). The straight line has slopand the circles from Eq.~(\ref{TK2}). The straight line has slope
$8/3$ and represents the prediction of Eq.~(\ref{TK}). Notice the
double-logarithmic scale.}
\end{figure}

\begin{figure}
\epsfysize=7.5cm
\epsffile{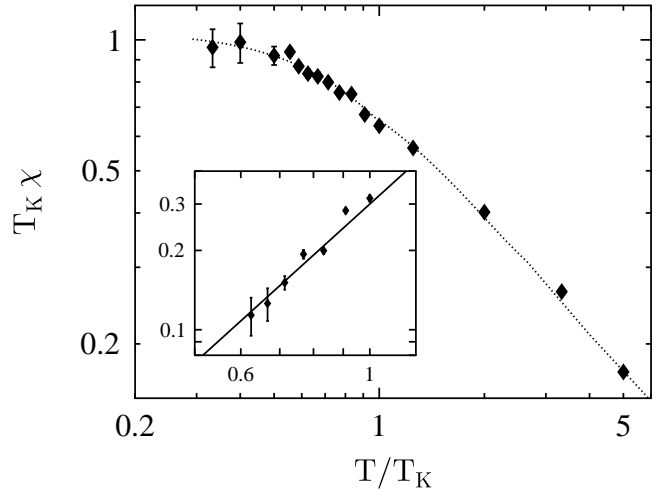}
\vspace{0.3cm}
\caption[]{\label{fig4}
Scaling curve for the temperature dependence of the impurity
susceptibility at $g_c=1/2$. The inset shows the same data for
$(\chi_0-\chi) T_K$ as a function of $T/T_K$ at low temperatures
again. The straight line in the
inset has slope $2$. Notice the double-logarithmic scales.
The dotted curve is a guide to the eye only.}
\end{figure}

\begin{figure}
\epsfysize=7.5cm
\epsffile{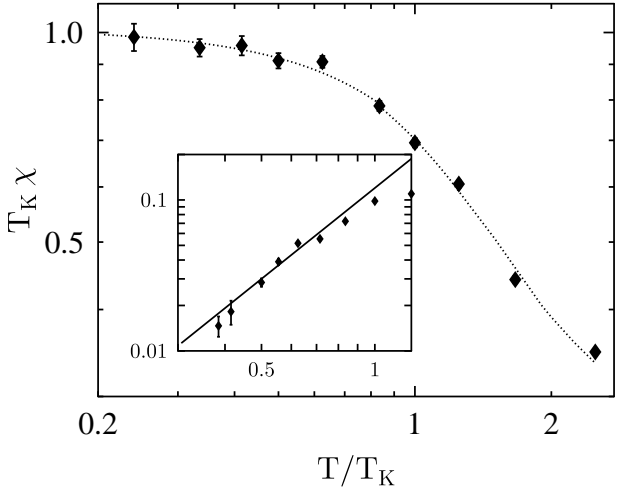}
\vspace{0.3cm}
\caption[]{\label{fig5}
Same as in Fig.~\ref{fig4} but for $g_c=1/4$. The
straight line in the inset has slope 2.}
\end{figure}

\begin{figure}
\epsfysize=7.5cm
\epsffile{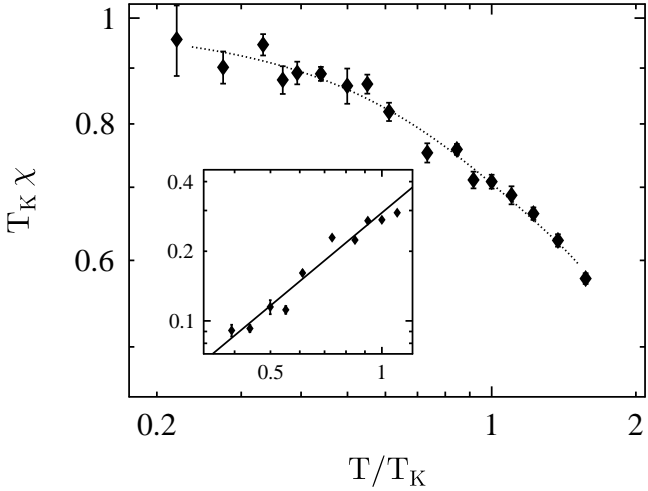}
\vspace{0.3cm}
\caption[]{\label{fig6}
Scaling curve as in Fig.~\ref{fig4} but for $g_c=3/4$. The
straight line in the inset has slope $4/3$.}
\end{figure}

The same critical behavior is found for $g_c=1/4$ as shown in
Fig.~\ref{fig5}. At low temperatures, the universal scaling curve displays
a $T^2$ behavior. The data shown here were obtained for $J/2 \pi v_F=0.1$, but
by virtue of scaling the same curve is found for other $J$ as well. From these
results one might be tempted to infer Fermi-liquid behavior for all interaction
strength parameters $g_c$. However, we find the 
Furusaki-Nagaosa $T^{1/g_c}$ law as soon
as $g_c > 1/2$, as shown in Fig.~\ref{fig6} for $g_c=3/4$.
The slope in the inset of Fig.~\ref{fig6} is $4/3$, in
accordance with the exponent found in Ref.~\onlinecite{furu}. 

\begin{figure}
\epsfysize=7.5cm
\epsffile{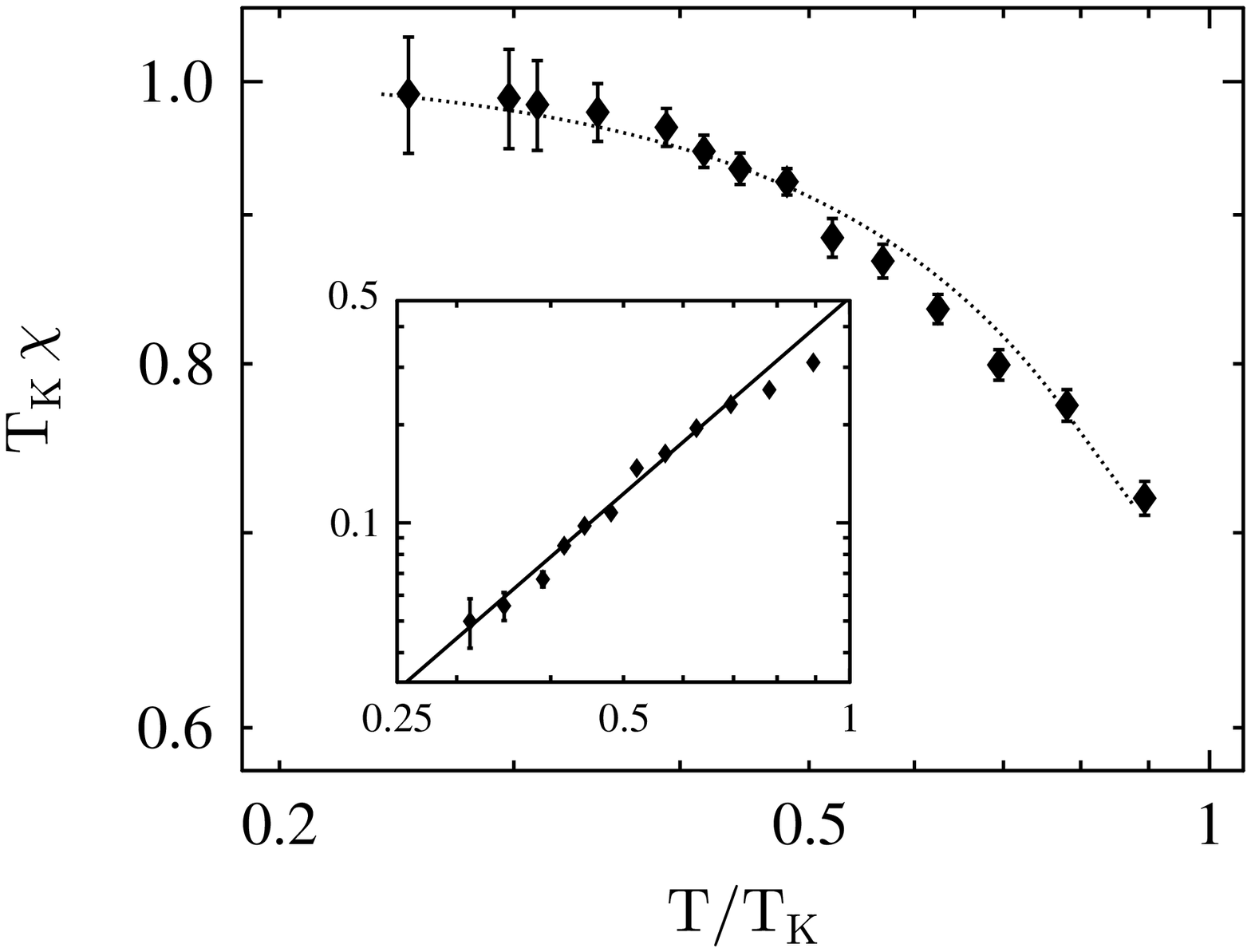}
\vspace{0.3cm}
\caption[]{\label{fig7}
Scaling curve as in Fig.~\ref{fig4} but for $g_c=0.001$. The
straight line in the inset has slope $2$.}
\end{figure}

In addition, we have analyzed the case of extremely strong interactions. This
formally corresponds to $g_c \to 0$. Assuming that this limit is
analytical, we can find the impurity susceptibility behavior from a study
of $g_c=0.001$, see Fig.~\ref{fig7}. Again, 
in accordance with our previous analysis, we find a
$T^2$ law for the low-temperature susceptibility. 

Let us now discuss these numerical results.
Our simulation data for the impurity 
susceptibility at $T \ll T_K$ obey the scaling form (\ref{TK1}) with
\begin{equation} \label{sug}
f(T/T_K) = 1 - c_1 (T/T_K)^2 - c_2 (T/T_K)^{1/g_c} 
+ \cdots \; .
\end{equation}
Hence there are two leading irrelevant scaling fields,\cite{cardy}
one describing Fermi liquid 
($\lambda_1$) and one describing non-Fermi liquid  
($\lambda_2$) behavior, where the latter one corresponds to the
Furusaki-Nagaosa prediction. For $g_c<1/2$, the 
Fermi-liquid term is more important and leads to the observed 
$T^2$ behavior at
low temperatures. In contrast, for $1/2<g_c<1$, 
the non-Fermi liquid behavior predicted in Ref.\onlinecite{furu}
dominates.

The scaling form (\ref{sug}) is consistent 
with the conformal-field theory analysis.\cite{henrik,henrik1} 
The operator $\widehat{O}_2$ conjugate to the scaling field $\lambda_2$
is produced by a composite
boundary operator in spin and charge sectors, while the operator
$\widehat{O}_1$ comes from a composite operator given by the products of
energy-momenta tensors in spin and charge sectors. As these are descendants of
the identity operator, their contribution 
to the susceptibility becomes linear in $\lambda_1$, i.e.,
 $c_1 \sim \lambda_1$. In contrast, the
Furusaki-Nagaosa term scaling like $T^{1/g_c}$ is quadratic in 
the corresponding scaling field,  $c_2\sim\lambda^2_2$.  
The amplitude of this contribution vanishes $\sim (1-g_c)$ as $g_c \to 1$, 
thereby reproducing the correct Fermi-liquid behavior of the conventional
Kondo effect for uncorrelated electrons.
Parenthetically, we note that the scaling field
$\lambda_2$ also produces a subleading $T^{1+1/g_c}$ law in the
impurity specific heat. 

To summarize, at $g_c<1/2$, the Fermi
liquid behavior will always dominate. 
However, at sufficiently low temperatures, the Furusaki-Nagaosa
 exponents can be
observed at $1/2<g_c<1$.  This finding is in conflict
with the recent numerical DMRG study by Wang,\cite{xwang} 
which reports Fermi-liquid behavior for a spin-$\frac12$
impurity coupled to a Hubbard chain. Notice that the interaction parameter
$g_c$ for the 1D Hubbard model away from half-filling
 is always within the bounds $1/2<g_c<1$.\cite{schulz96} 
Most likely the discrepancy is caused by finite-size effects 
due to the short chain lengths used in Ref.~\onlinecite{xwang}.
The more complicated outcome (\ref{sug}) also shows that the simplified model
by Schiller and Ingersent\cite{schiller} does not capture all essentials of
the Kondo effect in a Luttinger liquid. 

Finally, let us discuss our data for extremely strong interactions,
$g_c \to 0$. For the clean case, it is well established that  the
Luttinger liquid model for $g_c\to 0$ is equivalent to the 
low-energy sector of the 1D Heisenberg spin chain.\cite{schulz96}
Assuming that this reasoning carries over if a magnetic impurity is present,
the $T^2$ scaling of the impurity susceptibility observed here
(see Fig.~\ref{fig7} for $g_c=0.001$) should also describe the
susceptibility of a spin-$\frac12$ impurity interacting with a
1D Heisenberg chain. One has to be careful to couple the impurity 
to just one site of the Heisenberg chain, otherwise an 
additional potential scattering contribution will be present
(see Refs.\onlinecite{eggert,clarke,zhang,schlottmann}
 and Sec.~IV).
A different result was reported very recently by Liu,\cite{liu} 
namely a $T^{5/2}$ scaling of the impurity susceptibility at low
temperatures. Unfortunately, the
reason for this discrepancy is not clear at the moment.

\section{Critical impurity dynamics with potential scattering}

In this section the influence of elastic potential scattering
on the critical properties of a spin-$\frac12$ impurity 
in a Luttinger liquid will be discussed. 
As already mentioned in the Introduction,   
for sufficiently strong interaction strength, $g_c<1/2$, 
and for strong enough potential scattering,
the system is expected to display physics familiar from the
two-channel Kondo model.\cite{fabrizio}

\begin{figure}
\epsfysize=7.5cm
\epsffile{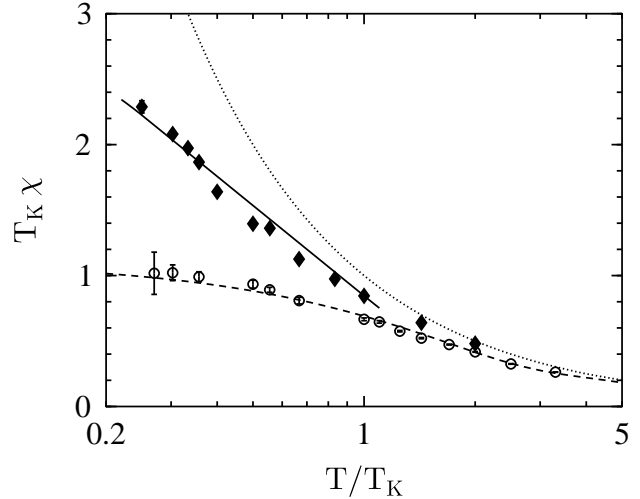}
\vspace{0.3cm}
\caption[]{\label{fig8}
Impurity susceptibility at $g_c=1/4$ in the presence of elastic
potential scattering, $2V/\pi v_F=0.2$, for
exchange coupling $J/2\pi v_F=0.08$.
The data points are given as filled diamonds. For comparison, the $V=0$ data
from Fig.~\ref{fig5} are shown as open circles.  The solid line
has slope $1/\pi \Gamma$ (see text), and the dotted curve gives has slope $1/\pi \Gamma$ (see text), and the dotted curve gives the
susceptibility (\ref{frs}) of a free spin.
The dashed curve is a guide to the eye only, and
$T_K$ is computed for $V=0$.
Notice the semi-logarithmic scales.}
\end{figure}

In Fig.~\ref{fig8}, data are shown 
for $g_c=1/4$.  At high temperatures, with 
or without elastic potential scattering, the Curie 
susceptibility of a free spin is always approached,
\begin{equation} \label{frs}
\chi_{\rm free}(T) = \beta/4 \;.
\end{equation}
However, while the impurity susceptibility displays a crossover  to the 
finite value $\chi_0=1/T_K$ at zero temperature
for vanishing elastic potential scattering strength ($V=0$), the 
behavior is drastically different if potential scattering is present (here,
$2V/\pi v_F=0.2$).  The impurity susceptibility does not saturate
but continues to increase without bound when lowering the temperature.
Since we have  semi-logarithmic scales in Fig.~\ref{fig8},
our data are accurately fitted by the susceptibility of the
two-channel Kondo model,\cite{kivelson}
\begin{equation}\label{twochan}
\chi(T) \simeq \frac{1}{\pi \Gamma} \ln \left(\frac{\Gamma}{T}\right)\;,
\end{equation}
where $\Gamma= J^2/2\pi^2 v_F^2$.  This value for $\Gamma$ 
(see Ref.\onlinecite{kivelson}) is 
indeed obtained from the slope of the solid line in Fig.~\ref{fig8}. 

\begin{figure}
\epsfysize=7.5cm
\epsffile{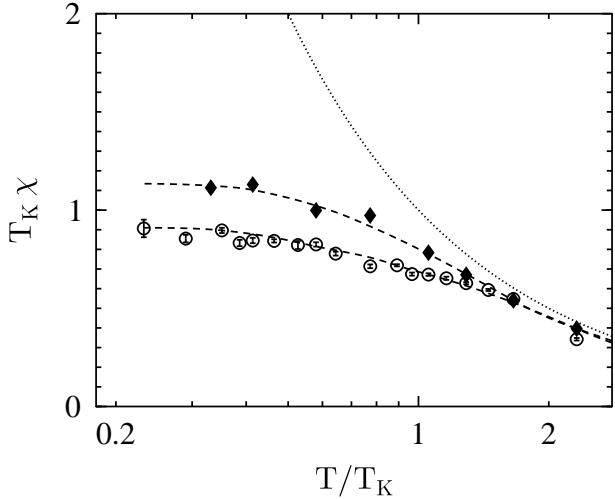}
\vspace{0.3cm}
\caption[]{\label{fig9}
Impurity susceptibility at $g_c=3/4$ in the presence of elastic
potential scattering, $2V/\pi v_F=0.3$, for
exchange coupling $J/2\pi v_F=0.1$.
Data points are given as filled diamonds. For comparison, the $V=0$ data
from Fig.~\ref{fig6} are shown as open circles.  The dotted curve gives the
susceptibility (\ref{frs}) of a free spin.  The dashed curves
are guides to the eye only,
and $T_K$ is computed for $V=0$.  Notice the semi-logarithmic scales.}
\end{figure}

The corresponding results for $g_c=3/4$ are shown in Fig.~\ref{fig9}.
The logarithmically divergent behavior in the presence
of potential scattering is not found anymore, and 
the low-temperature impurity susceptibility saturates at a 
finite value $\chi^V_0$.  Since $\chi_0^V> \chi_0$, we
expect from Eq.~(\ref{TK2}) that all effects of potential scattering
can be incorporated by a renormalization of the Kondo temperature $T_K$
to smaller values. Within  statistical error bars, 
the data for $2V/\pi v_F=0.3$ shown 
in Fig.~\ref{fig9} can indeed be scaled onto the 
$V=0$ data, and the scaling function $f$ holds even in the presence
of elastic potential scattering.  Clearly, this 
finding is in sharp contrast to the case of strong 
interactions, $g_c=1/4$, where potential 
scattering drastically changes the temperature
dependence of the impurity susceptibility.
  
From these data and the arguments of Ref.\onlinecite{fabrizio}, we then
expect two-channel Kondo behavior and hence a logarithmically
divergent susceptibility for all $g_c<1/2$.  An important
special case of this general result is recovered for $g_c\to 0$ 
which corresponds to
the 1D Heisenberg chain. Using numerical methods, 
bosonization and CFT techniques, Eggert and Affleck\cite{eggert}
and Clarke {\em et al.}\cite{clarke} have shown that
a spin-$\frac12$ impurity in a Heisenberg chain exhibits a
logarithmically divergent impurity susceptibility.
Due to the specific coupling of the impurity to the 1D spin
chain in these studies, an additional elastic potential scattering was 
present besides the usual Kondo exchange coupling term.

\section{Conclusions} 

In this paper the critical behavior of a spin-$\frac12$ impurity in
a correlated one-dimensional metal (Luttinger liquid)
has been investigated numerically.
To circumvent finite-size restrictions, we have developed
and applied a quantum Monte Carlo algorithm which allows to 
determine any finite-temperature equilibrium  quantity of interest.
Here we have focused on the impurity susceptibility $\chi$ with
particular emphasis on the low-temperature behavior well below the Kondo
temperature. Let us briefly summarize the main findings emerging from
our numerically exact analysis.

If elastic potential scattering is ignored,  the impurity susceptibility
shows the scaling behavior $T_K \chi(T) = f(T/T_K)$ with a distinct
universal scaling function $f$ for each dimensionless 
interaction strength $g_c$.
It may be worth mentioning that 
scaling holds even outside the
asymptotic low-temperature regime $T\ll T_K$. 
Within error bars, all data can be scaled onto universal
scaling functions as long as $T\ll \omega_c$, where $\omega_c$
is the bandwidth. 
Matching $\chi(T)$ curves for different $J$ (but at a given
$g_c$) onto a scaling curve also yields
the correct power-law dependence of the Kondo 
temperature, $T_K\sim J^{2/(1-g_c)}$,  which was first given in
Ref.\onlinecite{toner}.
 
We have then used our algorithm to determine the critical behavior
for $T\ll T_K$. Generally one finds power laws
$\chi(T)\sim (T/T_K)^\eta$ with some exponent
$\eta$.  At $g_c=1/4$ and $g_c=1/2$, we find $\eta=2$, but 
at $g_c=3/4$, a different exponent $\eta=4/3$ is obtained.
Our data are consistent with the simultaneous existence of
two leading irrelevant operators, one describing Fermi-liquid
behavior ($\eta=2$), the other describing the Furusaki-Nagaosa\cite{furu}
anomalous exponent $\eta=1/g_c$.
At $g_c<1/2$, the Fermi-liquid behavior is dominant, but 
at $1/2<g_c<1$, one can indeed observe the $g_c$-dependent exponents.
These findings resolve the recent 
controversy\cite{henrik,schiller,xwang,luyu,wang}
about the low-temperature criticality 
of the Kondo effect in a Luttinger liquid.

We have also studied the effects of elastic
potential scattering using our numerical approach.  As predicted
by Fabrizio and Gogolin,\cite{fabrizio} for sufficiently strong
Coulomb interaction strength, $g_c<1/2$, the impurity susceptibility
exhibits a logarithmically divergent behavior. The $\chi \sim \ln(1/T)$
scaling is a manifestation of two-channel Kondo physics caused
by the effectively open boundary at the impurity site. 
In contrast, for $1/2<g_c<1$, the susceptibility saturates
to a finite value at zero temperature and potential scattering
does not modify the critical behavior.

To conclude, we have numerically examined the critical scaling properties
of the Kondo effect in a Luttinger liquid. 
An interesting question which has not yet been studied in detail 
is related to universality in the presence of potential
scattering, e.g., the existence of universal scaling
functions for the impurity susceptibility.
Future applications of our Monte Carlo algorithm might also deal
with the case of more than one impurity, or with a systematic
study of other quantities like the impurity specific heat.

\acknowledgments

The authors would like to thank Henrik Johannesson for enlightening
discussions on the implications of conformal
field theory and acknowledge helpful conversations with Akira Furusaki,
Sasha Gogolin, Hermann Grabert, Herbert Schoeller and Johannes Voit.
This work has been supported by the Deutsche Forschungsgemeinschaft
(Bonn).

\end{document}